\documentclass[a4paper]{article}

\usepackage[utf8]{inputenc}
\usepackage[T1]{fontenc}

\usepackage{amsmath}
\usepackage{amssymb}
\usepackage{graphicx}
\usepackage[left=1in, right=1in, top=0.8in, bottom=1.2in]{geometry}
\usepackage{hyperref}
\usepackage[super]{nth}
\usepackage{siunitx}

\title{Robust direct acoustic impedance control using two microphones for mixed feedforward-feedback controller}
\author{
    Maxime Volery\thanks{Signal Processing Laboratory 2, EPFL, 1015 Lausanne, Switzerland}~\thanks{Corresponding author: maxime.volery@epfl.ch},
    Xinxin Guo\footnotemark[1],
    and Hervé Lissek\footnotemark[1]}

\DeclareMathOperator{\diag}{diag}

\begin{document}
\maketitle

\paragraph{Abstract}
    This paper presents an acoustic impedance control architecture for an electroacoustic absorber combining both feedforward and feedback microphone-based strategies on a current-driven loudspeaker.
    Feedforward systems enable good performance for direct impedance control.
    However, inaccuracies in the required actuator model can lead to a loss of passivity, which can cause unstable behaviour.
    The feedback contribution allows the absorber to better handle model errors and still achieve an accurate impedance, preserving passivity.
    Numerical and experimental studies were conducted to compare this new architecture against a state-of-the-art feedforward control method.


\section{Introduction}
Electroacoustic absorption consists in controlling the acoustic impedance presented by an electroacoustic actuator, typically an electrodynamic loudspeaker \cite{Olson53}.
The control of this impedance can be done passively, by loading the voice coil of the loudspeaker with an appropriate electrical impedance \cite{Fleming07,Boulandet16}, or actively, using one or more sensors controlling the voltage or current applied to the actuator.
Active electroacoustic absorbers have a wide range of applications, spanning from room acoustics \cite{Rivet17} to aircraft engine noise reduction \cite{Enoval18} thanks to their advantage of being tuneable, broadband and of sub-wavelength dimensions.
Most of the state-of-the art active absorber designs are either not tuneable, such as in the hybrid passive/active absorption concept \cite{Galland04} or require both a pressure and velocity sensor for a feedback implementation.
The sensing of the velocity can, for instance, be achieved using an accelerometer placed on the loudspeaker cone \cite{Cox04} (not acceptable for small loudspeakers), two closely placed microphones \cite{Orduna92} (not practical because upstream from the impedance plane) or a Wheatstone bridge \cite{Meynial99} (requires fine resistors and inductance tuning).

However, should the model of the actuator be known, a feedforward architecture \cite{Rivet17} can be used where only a single sensor is needed.
Also, thanks to the model inversion, direct impedance control can be achieved accurately, whereas other methods only approach the target impedance.
Nevertheless, due to some inevitable inaccuracies in the estimation of the model parameters and the delay of the numerical controller, a mismatch between the target impedance and the achieved one will eventually occur.
This mismatch can cause a loss of acoustic passivity of the absorber, meaning that it is injecting energy into the acoustic environment instead of absorbing it.
Such behaviour is unwelcome, even if it occurs outside of the frequency band of interest, because it can result in an unstable positive acoustic feedback.
In other words, if at a given frequency, the absorber injects more energy than the acoustic environment dissipates, energy will build-up, leading to an instability \cite{DeBono22}.

Combining both a feed-forward and a feedback loop can help reduce the inaccuracies while keeping the same performances, enabling a better fit with the analytical target impedance.
The membrane velocity estimation needed for the feedback implementation can be obtained via a microphone placed inside the cavity of the loudspeaker \cite{Guo20, Guo22}.
Indeed, for wavelengths smaller than the cabinet dimensions, the acoustic pressure behind the actuator is proportional to its membrane displacement and can be used to control it.
With this configuration, the size and complexity of the proposed mixed feedforward-feedback strategy does not fundamentally change from the former feedforward-only architecture and can be directly compared.

This paper is organized as follows.
In section \ref{sec:02_absorber_design}, a model of the electrodynamic loudspeaker is introduced before the description of the two-input control architecture.
Section \ref{sec:03_numerical_analysis} presents a Monte-Carlo analysis of the sensitivity of the achieved absorption to the model estimation errors.
Experimental validation of the proposed architecture is given in section \ref{sec:04_experimental_results} for three different control configurations, and section \ref{sec:05_conclusion} provides conclusion and opens some future perspectives for the presented concept.

\section{Robust electroacoustic absorber design}
\label{sec:02_absorber_design}
\subsection{Model of the electrodynamic loudspeaker}
An electrodynamic loudspeaker can be modelled as a mass-spring-damper system, of mass $M_{ms}$, mechanical compliance $C_{ms}$ and mechanical resistance $R_{ms}$
It is thus a second order resonator \cite{Rossi07}.
Three forces act on its membrane: the pressure in front of the membrane $p_f$, the pressure behind the membrane $p_b$ and the Lorentz force due to the current $i$ flowing in the voice coil.
When mounted on an enclosure, the contribution from the rear pressure can be modelled as a specific compliance $C_{sb}$ for wavelengths smaller than the cabinet dimensions.
This compliance is the ratio between membrane displacement and the pressure in the cavity, and is linked to the volume of the cavity $V_b$ as follows:
\begin{equation}
    C_{sb}=\frac{V_b}{\rho_0c_0^2S_d} ,
\end{equation}
where $\rho_0$ is the mass density of air, $c_0$ the speed of sound in the air, and $S_d$ the effective piston area of the loudspeaker.
The membrane motion is described by Newton's second law of motion
\begin{equation}
    M_{ms}\frac{dv(t)}{dt} = S_dp_f(t) - R_{ms}v(t) - \underbrace{\left(\frac{1}{C_{ms}} + \frac{S_d}{C_{sb}}\right)}_{1/C_{mc}}\int_0^tv(t)dt - Bli(t) ,
    \label{eq:newton_TD}
\end{equation}
where $v$ is the membrane inwards velocity, $Bl$ the coil force factor, and $C_{mc}$ the combined mechanical compliance of the loudspeaker and the cabinet.
The pressure in the cabinet $p_b$ is directly proportional to the membrane displacement
\begin{equation}
    p_b(t) = \frac{1}{C_{sb}}\int_0^tv(t)dt .
    \label{eq:pb_TD}
\end{equation}

In the Laplace domain, with Laplace variable $s$, equations \eqref{eq:newton_TD} and \eqref{eq:pb_TD} are written
\begin{equation}
    p_f(s) = Z_{ss}(s)v(s) + Fi(s)
    \label{eq:newton_FD}
\end{equation}
and
\begin{equation}
    p_b(s) = \frac{v(s)}{sC_{sb}} ,
    \label{eq:pb_FD}
\end{equation}
where
\begin{equation}
    F = \frac{Bl}{S_d} ,
\end{equation}
\begin{equation}
    Z_{ss}(s) = R_{ss}\frac{s^2 + s\omega_0/Q_{ms} + \omega_0^2}{s\omega_0/Q_{ms}}
    \label{eq:zm_def}
\end{equation}
is the specific impedance of the loudspeaker, $R_{ss}=R_{ms}/S_d$ its specific resistance, $\omega_0=1/\sqrt{M_{ms}C_{mc}}$ its natural resonance angular frequency and $Q_{ms}=R_{ms}^{-1}\sqrt{M_{ms}/C_{mc}}$ its (passive) quality factor.
From the representation of the impedance of equation \eqref{eq:zm_def}, it is straightforward to notice that the passive loudspeaker ($i=0$) mounted on a cabinet is indeed a second order resonator.

Because an accurate model of the electrical impedance of the loudspeaker is complex to develop and to estimate \cite{Leach02,Thorborg08}, and that the electrical force applied on the membrane is directly proportional to the current flowing in the coil, as shown in equation \eqref{eq:newton_FD}, it is interesting to drive the loudspeaker using a current source rather than a voltage source, as has been done in \cite{Rivet17}.
In the following, the loudspeaker is driven in current.
An implementation of such a current source is given in appendix \ref{sec:06_current_source}.

\subsection{Formulation of the Two-Input Single-Output controller}
Direct impedance control allows to reach a desired target impedance $Z_{st}(s)$ on the membrane of the loudspeaker instead of the passive one $Z_{ss}(s)$.
A feedforward-controller \cite{Rivet17} measures the pressure in front of the membrane and relies on the model of the actuator to find the current to inject in the voice coil to get the appropriate membrane velocity such that the desired target impedance is met.
It is therefore capable of reaching a wide range of target impedances.
However, this also implies that an accurate model of the loudspeaker must be given to the controller, and that any inaccuracy in this model can have an important impact on the obtained results (i.e., the achieved impedance will deviate from the target one).
Adding a feedback loop along with the feedforward architecture can help reduce this problem.
To implement feedback on top of the feedforward architecture, a measure of the velocity of the membrane is needed in addition to the pressure in front of it.
This can be achieved by sensing the pressure in the cavity closing the rear face of the actuator because the pressure in it is proportional to the displacement of the membrane at low frequencies, as shown in equation \eqref{eq:pb_FD}.

\begin{figure}
    \centering
    \includegraphics{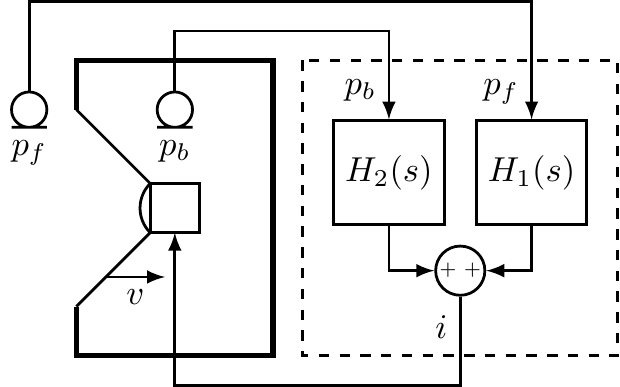}
    \caption{Controlled absorber. The two-input controller is depicted on the right in the dashed rectangle.}
    \label{fig:schematic}
\end{figure}

\begin{figure}
    \centering
    \includegraphics{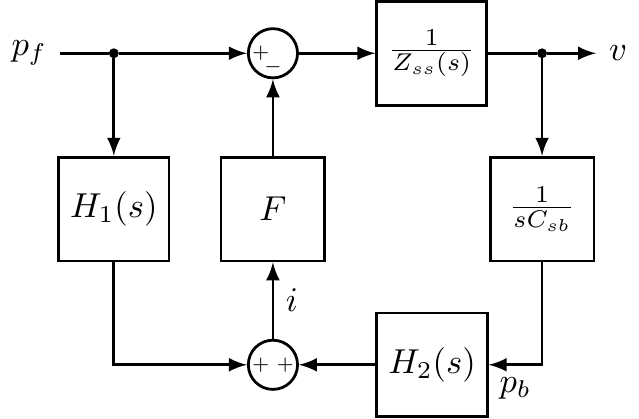}
    \caption{Block diagram of the mixed feedforward-feedback controlled absorber}
    \label{fig:block_diagram}
\end{figure}

It appears now that the controller has two inputs: the pressure in front of the membrane $p_f$ and the pressure behind it $p_b$ and has a single output: the current $i$ injected in the moving coil of the loudspeaker.
This output current can therefore be expressed as 
\begin{equation}
    i(s) = H_1(s)p_f(s) + H_2(s)p_b(s) ,
    \label{eqn:MISO_def}
\end{equation}
where both $H_1$ and $H_2$ are linear time-invariant systems.
An illustration of such a controller is shown in Figure \ref{fig:schematic}, and its detailed block diagram in Figure \ref{fig:block_diagram}. In the latter, it is clearly visible that $H_1(s)$ is the feedforward part of the controller and $H_2(s)$ the feedback part.

In order to achieve a target impedance $Z_{st}(s)$, it follows from equations \eqref{eq:newton_FD}, \eqref{eq:pb_FD} and \eqref{eqn:MISO_def} that $H_1$ and $H_2$ must satisfy the relation
\begin{equation}
    H_1(s) + \frac{H_2(s)}{sC_{sb}Z_{st}(s)} = \frac{1}{F}\left(1 - \frac{Z_{ss}(s)}{Z_{st}(s)}\right) .
    \label{eq:H1_H2_link}
\end{equation}

There is an infinite number of realizations that satisfy equation \eqref{eq:H1_H2_link}, but feedback from the membrane velocity is desired.
This feedback in velocity $G(s)$ is the combination of the controller $H_2$, the compliance of the enclosure and the force factor.
And because the modelling of the box as a constant compliance is only valid for wavelengths smaller than the dimension of the box, $G(s)$ should have a low-pass behaviour.
A first order low-pass filter is chosen for $G(s)$ such that the controller is of the smallest degree possible:
\begin{equation}
    G(s) = \frac{FH_2(s)}{sC_{sb}} = \rho_0c_0k_g\frac{\omega_g}{s+\omega_g} ,
    \label{eq:G_def}
\end{equation}
where $k_g\geq0$ is a dimensionless tuneable feedback gain and $\omega_g$ is the cut-off angular frequency of the low-pass filter $G(s)$. The two control transfer functions are thus
\begin{equation}
    H_1(s) = \frac{1}{F}\left(1 - \frac{Z_{ss}(s) + G(s)}{Z_{st}(s)}\right)
    \label{eq:proper_H1}
\end{equation}
and
\begin{equation}
    H_2(s) = \frac{sC_{sb}G(s)}{F} .
    \label{eq:proper_H2}
\end{equation}

In equations \eqref{eq:proper_H1} and \eqref{eq:proper_H2}, it can be observed that the controller is proper, and that by setting $G=0$, only $H_1(s)$ is left, and is equal to the state-of-the art feedforward controller from \cite{Rivet17} without any feedback.
Furthermore, equations \eqref{eq:proper_H1} and \eqref{eq:proper_H2} can also be interpreted as the superposition of the pure feedforward implementation and a pure feedback implementation where the error between target velocity and achieved velocity is fed as a current to the loudspeaker with feedback gain $G(s)$, as in \cite{Volery20}.

However, not any arbitrary impedance can be achieved: to avoid divergence of the control transfer function $H_1(s)$ for low and high frequencies, the asymptotes of the target impedance should behave as a compliance for low frequency, and a mass for high frequencies, as it is the case for the passive impedance.
In this article, the considered target impedance is a multi-degree-of-freedom resonator, which is the result of $N$ second order resonators connected in parallel, as used in \cite{Rivet17_2}
\begin{equation}
    Z_{st}(s) = \left( \sum_{n=1}^{N}
        \frac{1}{R_{st,n}}\,\frac{s\omega_{t,n}/Q_{t,n}}{s^2 + s\omega_{t,n}/Q_{t,n} + \omega_{t,n}^2}
    \right)^{-1} ,
    \label{eq:zt_def}
\end{equation}
where $R_{st,n}$, $\omega_{t,n}$ and $Q_{t,n}$ are respectively the specific resistance, the resonance angular frequency and the quality factor of the $n^{\text{th}}$ resonator.
Different realizations of the target impedance could also be considered, but the following derivation will consider the form of equation \eqref{eq:zt_def} without loss of generality.

\subsection{Proof of stability}
A pole analysis of the feedback loop created by $H_2(s)$ is required to show the stability properties of the absorber.
Each transfer functions $H_1(s)$ and $H_2(s)$ are individually (open loop) proper and stable.
There is one feed-forward loop, which is stable if its components are stable, and a feedback loop which is stable if the real part of all its poles is negative.
These poles are the solutions of
\begin{equation}
    \frac{1}{T(s)} = G(s) + Z_{ss}(s) = 0 ,
\end{equation}
where $T(s)$ is the closed loop transfer function between $(1 - FH_1)p_f$ and $v$.
This is equivalent to solving
\begin{equation}
    s^3 + as^2 + bs + c = 0 ,
    \label{eq:poles_eqn}
\end{equation}
where
\begin{equation}
    a = \frac{\omega_0}{Q_{ms}} + \omega_g ,
\end{equation}
\begin{equation}
    b = \omega_0^2 + \frac{\omega_0\omega_g}{Q_{ms}}\left(\frac{\rho_0c_0k_g}{R_{ss}} + 1\right)
\end{equation}
and
\begin{equation}
    c = \omega_0^2\omega_g, 
\end{equation}
and it is interesting to notice that equation \eqref{eq:poles_eqn} does not depend on the target impedance.
The closed loop $T(s)$ is stable if and only if the Hurwitz matrix \begin{equation}
    \mathcal{H} = \begin{bmatrix}
       a & c & 0 \\
       1 & b & 0 \\
       0 & a & c
    \end{bmatrix}
\end{equation}
corresponding to the polynomial of equation \eqref{eq:poles_eqn} has all its three leading principal minors which are positive \cite{Hurwitz95}:
\begin{equation}
    a > 0 ,
\end{equation}
\begin{equation}
    \begin{vmatrix}
        a & c \\ 1 & b
    \end{vmatrix} = ab - c > 0
\end{equation}
and
\begin{equation}
    \begin{vmatrix}
        a & c & 0 \\
        1 & b & 0 \\
        0 & a & c
    \end{vmatrix} = c\left(ab - c\right) > 0 .
\end{equation}
This means that $k_g$ must satisfy
\begin{equation}
    k_g > -\frac{R_{ss}}{\rho_0c_0}\left(1 + \frac{Q_{ms}\left(\omega_0/\omega_g\right)^2}{Q_{ms} + \omega_0/\omega_g}\right) ,
\end{equation}
which is always true for nonnegative values of $k_g$.

\subsection{Sensitivity to parameter variations}
To analyse the robustness of the proposed method to parameter estimation accuracy, the sensitivity functions of the achieved impedance are calculated.
When the estimated values $\hat{Z}_{ss}$, $\hat{F}$ and $\hat{C}_{sb}$ of the parameters $Z_{ss}$, $F$ and $C_{sb}$ respectively are used in the controller transfer functions from equations \eqref{eq:proper_H1} and \eqref{eq:proper_H2}, the achieved impedance is
\begin{equation}
    Z_{sa} = Z_{st}\frac{G(s)\hat{C}_{sb}/C_{sb} + Z_{ss}(s)\hat{F}/F}{G(s) + \hat{Z}_{ss}(s) + Z_{st}(s)\left(\hat{F}/F - 1\right)} .
    \label{eq:zsa}
\end{equation}
The sensitivity function of this achieved impedance with respect to a parameter $x$ is defined as the ratio between the percentage of change in the achieved impedance $Z_{sa}$ to the percentage of change in the parameter $x$ \cite{Shinners98}:
\begin{equation}
    S_x(s) =  \frac{\partial Z_{sa}}{\partial x} \frac{x}{Z_{sa}} .
\end{equation}
which results in
\begin{equation}
    S_{\hat{Z}_{ss}}(s) = -\left( 1 + \frac{G + \left(\hat{F}/F - 1\right)Z_{st}}{\hat{Z}_{ss}} \right)^{-1} ,
\end{equation}
\begin{equation}
    S_{\hat{F}}(s) = \left( 1 + \frac{\hat{C}_{sb}FG}{C_{sb}\hat{F}Z_{ss}} \right)^{-1}
    -  \left( 1 + \frac{F\left( G + \hat{Z}_{ss} - Z_{st}\right)}{\hat{F}Z_{st}} \right)^{-1}
\end{equation}
and
\begin{equation}
    S_{\hat{C}_{sb}}(s) = \left(1 + \frac{C_{sb}\hat{F}Z_{ss}}{\hat{C}_{sb}FG}\right)^{-1} ,
\end{equation}
for parameters $\hat{Z}_{ss}$, $\hat{F}$ and $\hat{C}_{sb}$ respectively.
The limit when $G(s) \rightarrow \infty$ of $S_{\hat{Z}_{ss}}(s)$, $S_{\hat{F}}(s)$ and $S_{\hat{C}_{sb}}(s)$ are respectively $0$, $0$ and $1$.
It can therefore be concluded that any variation in the estimation $\hat{Z}_{ss}$ and $\hat{F}$ will be less significant when the magnitude of $G(s)$ is larger.
This is however not true for $\hat{C}_{sb}$, for which the error on the achieved impedance becomes proportional to the error in $\hat{C}_{sb}$ when the magnitude of $G(s)$ is large.

\section{Numerical sensitivity analysis}
\label{sec:03_numerical_analysis}
In this section, a numerical sensitivity analysis is presented for three different control targets: a single-degree-of-freedom resonator whose resonance is shifted with respect to the passive one, a broadband absorption centred at the passive resonance and a two-degree-of-freedom impedance with two distinct shifted resonances.
The target impedances and the control parameters are defined according to equation \eqref{eq:zt_def} and are reported for each case in Table \ref{tbl:ctrl_parameters}.

\begin{table*}
    \centering
    \caption{Target impedances and control parameters for the three considered configurations}
    \label{tbl:ctrl_parameters}
    \begin{tabular}{lcccc}
        \hline
        \textbf{Parameter}          & \textbf{Symbol}   & \textbf{1 DOF}    & \textbf{Broadband}    & \textbf{2 DOF} \\
        \hline
        Specific resistance         & $R_{st}$          & $\rho_0c_0$       & $\rho_0c_0$           & $\rho_0c_0$ and $\rho_0c_0$ \\
        Resonance frequency         & $\omega_t/(2\pi)$ & \SI{400}{\hertz}  & \SI{200}{\hertz}      & \SIlist{100;400}{\hertz} \\
        Quality factor              & $Q_t$             & 7                 & 0.25                  & 7 and 7 \\
        Feedback gain               & $k_g$             & 4                 & 4                     & 4 \\
        Feedback cut-off frequency  & $w_g/(2\pi)$      & \SI{500}{\hertz}  & \SI{500}{\hertz}      & \SI{500}{\hertz} \\
        \hline
    \end{tabular}
\end{table*}

\begin{figure}
    \centering
    \includegraphics{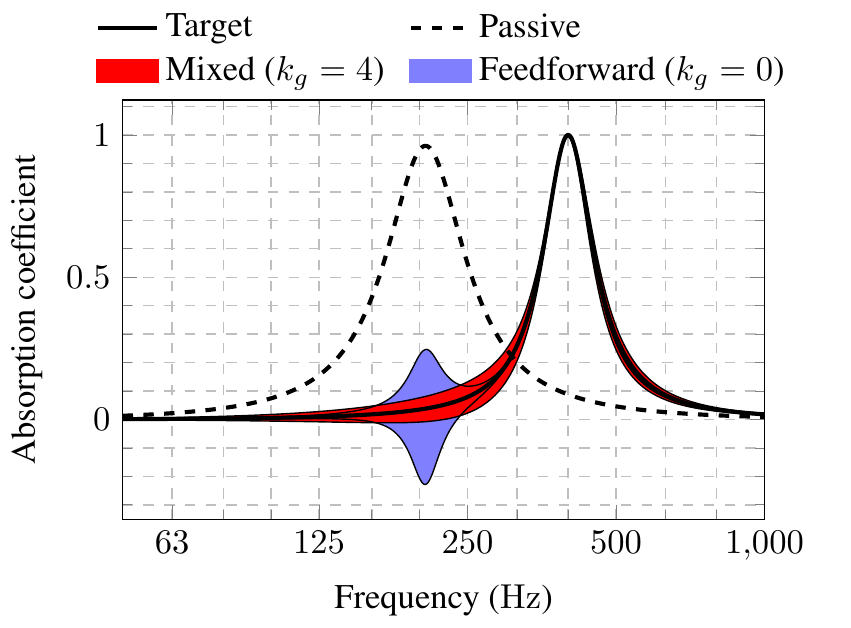}
    \caption{First and third quartiles of the achieved absorption for the single-degree-of-freedom absorber with $10^5$ random relative errors of 5\% standard deviation on the five estimated parameters}
    \label{fig:monte_carlo_1DOF}
\end{figure}

\begin{figure}
    \centering
    \includegraphics{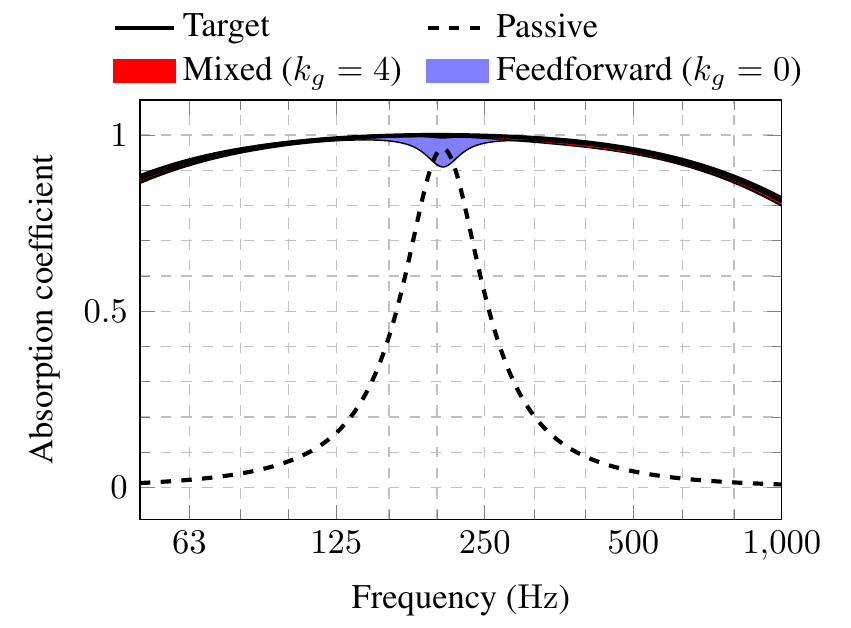}
    \caption{First and third quartiles of the achieved absorption for the broadband absorber with $10^5$ random relative errors of 5\% standard deviation on the five estimated parameters}
    \label{fig:monte_carlo_broadband}
\end{figure}

\begin{figure}
    \centering
    \includegraphics{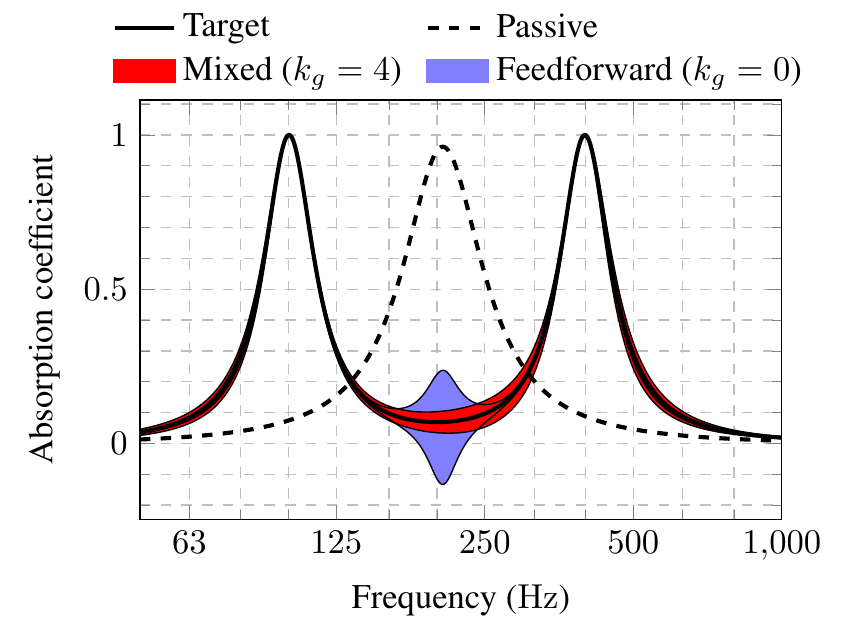}
    \caption{First and third quartiles of the achieved absorption for the two-degree-of-freedom absorber with $10^5$ random relative errors of 5\% standard deviation on the five estimated parameters}
    \label{fig:monte_carlo_2DOF}
\end{figure}

The numerical sensitivity analysis consists in evaluating the achieved normal incidence absorption coefficient $\alpha_a$ $10^5$ times, with random Gaussian deviations of 5\% on the estimated parameters $\hat{R}_{ss}$, $\hat{\omega}_0$, $\hat{Q}_{ms}$, $\hat{F}$ and $\hat{C}_{sb}$.
This absorption coefficient is defined as the ratio between absorbed and incident power.
It lies between 0 and 1 for acoustically passive systems, whereas it is smaller than one if the system is acoustically active (for which energy is injected in the acoustic domain instead of being absorbed).
It is calculated from the achieved impedance $Z_{sa}(s)$ as
\begin{equation}
    \alpha_a(s) = 1 - \left|\frac{Z_{sa}(s) - \rho_0c_0}{Z_{sa}(s) + \rho_0c_0}\right|^2 ,
\end{equation}
where $Z_{sa}$, the achieved impedance is evaluated according to equation \eqref{eq:zsa}.
At every simulated frequency, the values of the first and the third quartiles of the absorption coefficient are reported in Figure \ref{fig:monte_carlo_1DOF}, Figure \ref{fig:monte_carlo_broadband} and Figure \ref{fig:monte_carlo_2DOF} for each considered target.
In these figures, it is observable that the absorption coefficient with only feedforward deviates further away from the target than with the mixed feedforward-feedback control.
It can even reach negative values around the passive resonance of the actuator.
With feedback however, it is much better controlled around this resonance, but at the price of lower accuracy for other frequencies.

Although the feedback does not bring much improvement for the broadband absorption shown in Figure \ref{fig:monte_carlo_broadband}, it does for the two other cases.
In an Ultra High Bypass Ratio aircraft engine application, the sound to absorb is typically tonal, and an absorber with multiples rays of absorptions would be convenient \cite{Enoval18, Salze19}.
Also, in this application, the optimal impedance would not be $\rho_0c_0$ but rather consists of a given resistive part and a reactive part, as explained in \cite{Tester73}, for which this new architecture can bring interesting improvements.

\section{Experimental results}
\label{sec:04_experimental_results}
\subsection{Experimental setup}
\begin{figure*}
    \centering
    \includegraphics{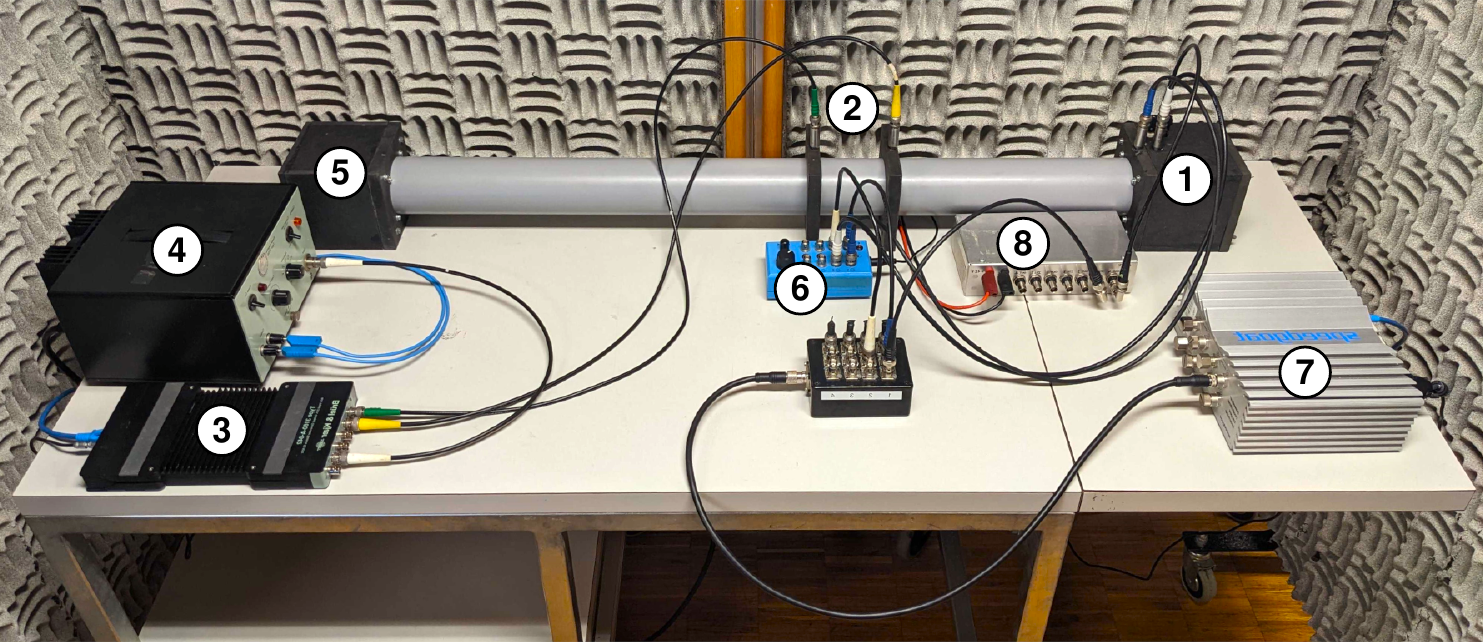}
    \caption{Experimental setup used to measure the impedance presented by the absorber. 1)~Electroacoustic resonator 2)~measurement microphones 3)~frequency analyser 4)~power amplifier 5)~sound source 6)~IEPE signal conditioner 7)~FPGA controller 8) current pump}
    \label{fig:06_experimental_setup_anotated}
\end{figure*}

\begin{figure*}
    \centering
    \includegraphics{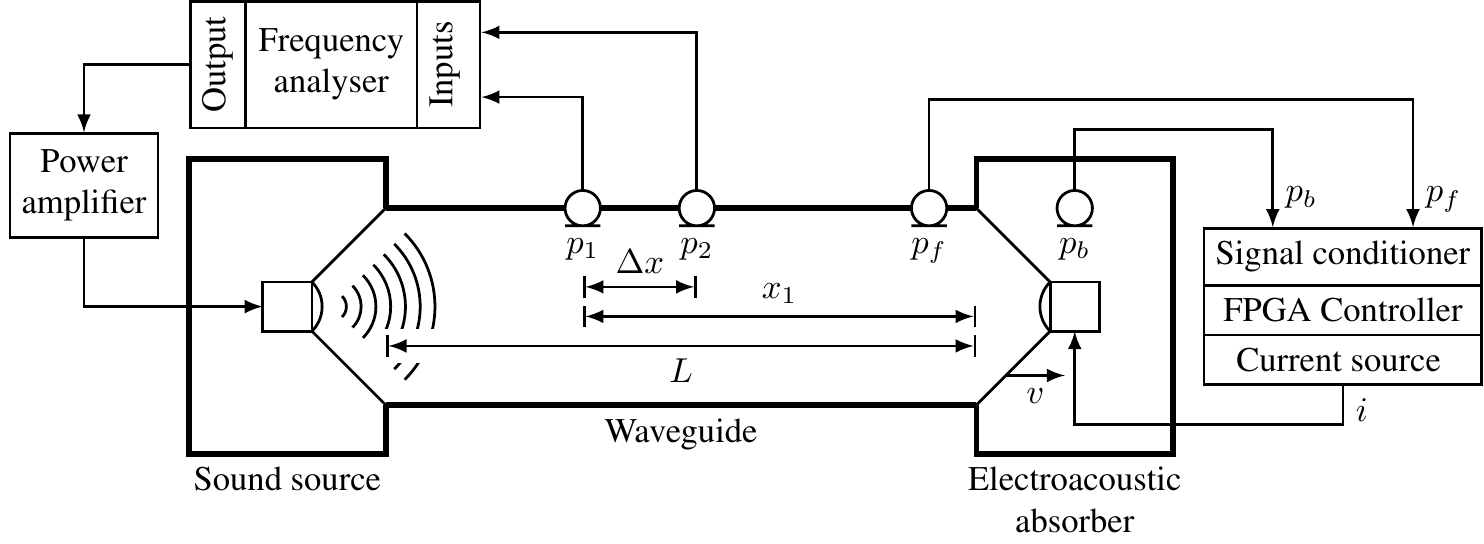}
    \caption{Schematic of the experimental setup used to measure the impedance presented by the absorber}
    \label{fig:experimental_setup}
\end{figure*}

The measurement setup used to experimentally assess this new control architecture is shown in Figure \ref{fig:06_experimental_setup_anotated}, and schematised in Figure \ref{fig:experimental_setup}.
The two microphones used to control the electroacoustic absorber are connected to the field-programmable gate array (FPGA) controller through a signal conditioner.
The digital filter running on the FPGA is the bilinear transform of equations \eqref{eq:proper_H1} and \eqref{eq:proper_H2} with a sampling frequency of \SI{50}{\kilo\hertz}.
For better numerical stability, the digital filter is realized as a cascade of second-order sections \cite{Mitra98}.
The output voltage of the controller is converted into a current by a home-made voltage-controlled current source whose schematic is described in appendix \ref{sec:06_current_source}.
A short study on the impact of the position of the rear microphone is available in appendix \ref{sec:06_microphone_placement}.

The achieved impedance presented by the absorber is measured using a Kundt's tube after ISO 10534-2 \cite{ISO_10534_2}.
A multichannel frequency analyser feeds white noise to the amplified external source during \SI{60}{\second} (resulting in a sound pressure level up to \SI{105}{\decibel} at the absorber position) while measuring the signals from the two measurement microphones $p_1$ and $p_2$.
From the transfer function $p_2(s)/p_1(s)$ and the waveguide dimensions $\Delta x$ and $x_1$, the reflection coefficient of the termination of the waveguide, and thus its impedance too, can be recovered \cite{ISO_10534_2}.
The estimation of the transfer function is done with a linear averaging of \SI{1}{\second} length Hann windows overlapping by 66.67\%, with a \SI{1}{\hertz} frequency resolution.
All the hardware equipment used is listed in Table \ref{tbl:experimental_setup}.

\begin{table}
    \centering
    \caption{Experimental setup equipment list}
    \label{tbl:experimental_setup}
    \begin{tabular}{ll}
        \hline
        \textbf{Equipment}      & \textbf{Model} \\
        \hline
        Microphone type         & PCB 130D20 \\
        IEPE signal conditioner & MMF M31 \\
        FPGA controller         & Speedgoat IO334 \\
        Frequency analyser      & Brüel \& Kjær type 3160 \\
        Power amplifier         & Brüel \& Kjær type 2706 \\
        Waveguide dimensions    & $\Delta x$: \SI{100}{\milli\meter}, $x_1$: \SI{420}{\milli\meter} \\
                                & $L$: \SI{970}{\milli\meter}, $\varnothing$: \SI{72}{\milli\meter} \\
        \hline
    \end{tabular}
\end{table}

\subsection{Transducer parameters identification}
\label{sec:experimental_setup}
To implement the filters from equations \eqref{eq:proper_H1} and \eqref{eq:proper_H2}, five parameters of the electrodynamic loudspeaker are needed: $R_{ss}$, $\omega_0$, $Q_{ms}$, $F$ and $C_{sb}$.
The estimation of the specific mass $M_{ss} = M_{ms}/S_d$, resistance $R_{ss}$ and stiffness $K_{sc}=1/(S_dC_{mc})$ are obtained by a polynomial fitting of the measured passive ($i=0$) impedance curve:
\begin{equation}
    \begin{bmatrix}
        \hat{M}_{ss} \\
        \hat{R}_{ss} \\
        \hat{K}_{sc}
    \end{bmatrix}
    = \begin{bmatrix}
        \mathbf{0} & \mathbf{1} & \mathbf{0} \\
        \boldsymbol\omega & \mathbf{0} & -\diag(\boldsymbol\omega)^{-1}\mathbf{1}
    \end{bmatrix}^+
    \begin{bmatrix}
        \Re\left\{Z_{ss}(j\boldsymbol\omega)\right\} \\ \Im\left\{Z_{ss}(j\boldsymbol\omega)\right\}
    \end{bmatrix},
    \label{eq:polyfit}
\end{equation}
where $+$ denotes the Moore-Penrose pseudo-inverse, $\boldsymbol\omega$ is a vector containing the measured angular frequencies, $Z_{ss}(s)$ is the measured specific impedance and $\mathbf{0}$ and $\mathbf{1}$ are vectors of respectively zeros or ones of the same size as $\boldsymbol\omega$.
The parameters $\omega_0$ and $Q_{ms}$ are straightforward to derive from the result of equation \eqref{eq:polyfit}.
Then, $F$ can be estimated as presented in \cite{DeBono21}, using the proportional controller $i=K_1p_f$:
\begin{equation}
    \hat{F} = \Re\left\lbrace\frac{1}{N}\sum_{n=1}^N \frac{1 - Z_{ss}(j\omega_n)/Z_1(j\omega_n)}{K_1}\right\rbrace ,
    \label{eq:blsd_estimate}
\end{equation}
where $Z_1(s)$ is the specific impedance measured with the constant feedforward controller of gain $K_1$ and $\omega_n$ is the $n^\text{th}$ element of $\boldsymbol\omega$.
Finally, the box specific compliance can be found using the proportional controller $i=K_2p_b$:
\begin{equation}
    \hat{C}_{sb} = \Re\left\lbrace\frac{1}{N}\sum_{n=1}^N \frac{\hat{F}K_2/(j\omega_n)}{Z_2(j\omega_n) - Z_{ss}(j\omega_n)}\right\rbrace ,
    \label{eq:cb_estimate}
\end{equation}
where $Z_2(s)$ is the specific impedance measured with the constant feedback controller of gain $K_2$.

All these measured parameters of the electrodynamic absorber are reported in Table \ref{tbl:ldsp_meas}.
The frequency band considered in equations \eqref{eq:polyfit}, \eqref{eq:blsd_estimate} and \eqref{eq:cb_estimate} is from \SI{170}{\hertz} to \SI{250}{\hertz} with steps of \SI{1}{\hertz}.
Note that these parameters describe the termination of the Kundt's tube.
To get the loudspeaker parameters, they must be scaled by $S_d/S_\text{duct}$, where $S_\text{duct}$ is the cross section of the duct.
However, this is not necessary if one is interested in controlling the impedance of the whole termination instead of only the loudspeaker.
Indeed, using the cross section $S_\text{duct}$ instead of $S_d$ is equivalent to a scaling of $v$, and thus a scaling of the impedances and the box compliance.
It therefore has no impact on the equations if all the measured impedances as well as the target one are considered with the same cross-section.
It is also interesting to notice that the calibration of the two control microphones is not necessary. Indeed, in both equations \eqref{eq:blsd_estimate} and \eqref{eq:cb_estimate} the errors in the microphone sensitivities are embedded in the estimation of $F$ and $C_{sb}$.

\begin{table}
    \centering
    \caption{Measured Thiele-Small parameters of the Monacor SPX-30M loudspeaker mounted on a cabinet}
    \label{tbl:ldsp_meas}
    \begin{tabular}{lcc}
        \hline
        \textbf{Parameter}      & \textbf{Symbol} & \textbf{Value} \\
        \hline
        Specific resistance     & $R_{ss}$      & 0.6734$\rho_0c_0$ \\
        Resonant frequency      & $w_0/(2\pi)$  & \SI{205.5}{\hertz} \\
        Mechanical Q factor     & $Q_{ms}$      & 5.466 \\
        Box spec. compliance    & $C_{sb}$      & \SI{1.808}{\micro\meter\per\pascal} \\
        Pressure factor         & $F$           & \SI{1.084}{\pascal\per\milli\ampere} \\
        Density of air          & $\rho_0$      & \SI{1.2}{\kilogram\per\meter\cubed} \\
        Speed of sound          & $c_0$         & \SI{343}{\meter\per\second} \\
        \hline
    \end{tabular}
\end{table}

\subsection{Impedance measurements}
\begin{figure}
    \centering
    \includegraphics{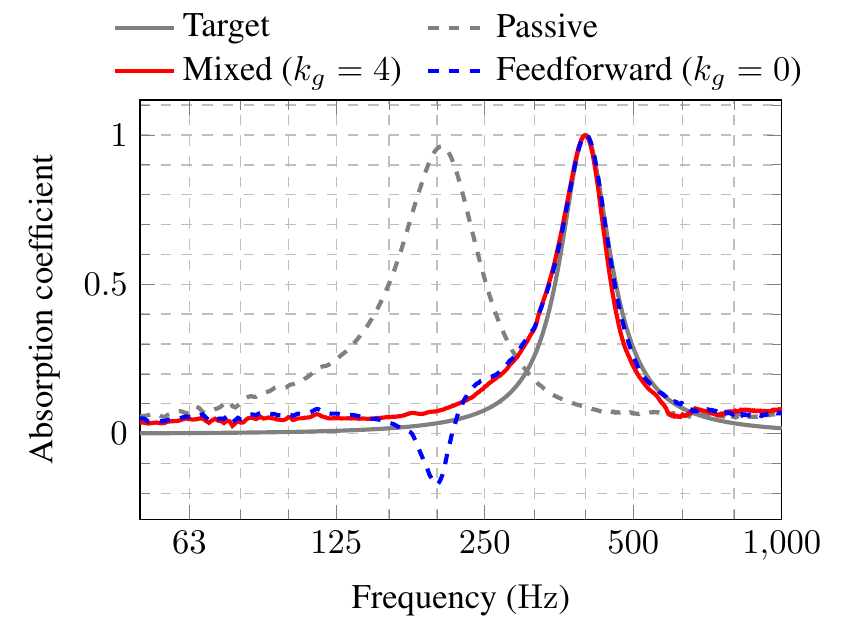}
    \caption{Experimentally obtained absorption coefficients for the single-degree-of-freedom absorber, with $\hat{F} = 0.95F$}
    \label{fig:experimental_1DOF_alpha}
\end{figure}

\begin{figure}
    \centering
    \includegraphics{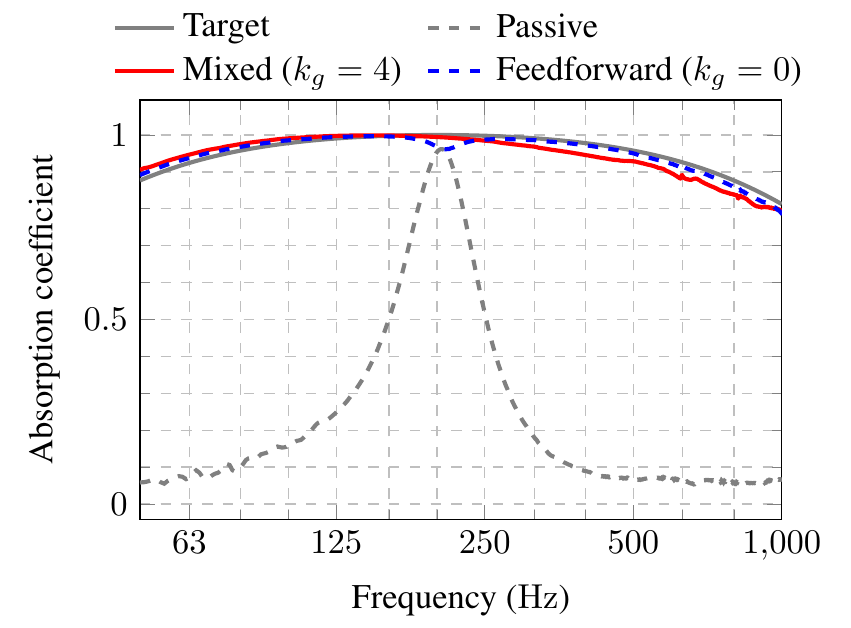}
    \caption{Experimentally obtained absorption coefficients for the broadband absorber, with $\hat{F} = 0.95F$}
    \label{fig:experimental_broadband_alpha}
\end{figure}

\begin{figure}
    \centering
    \includegraphics{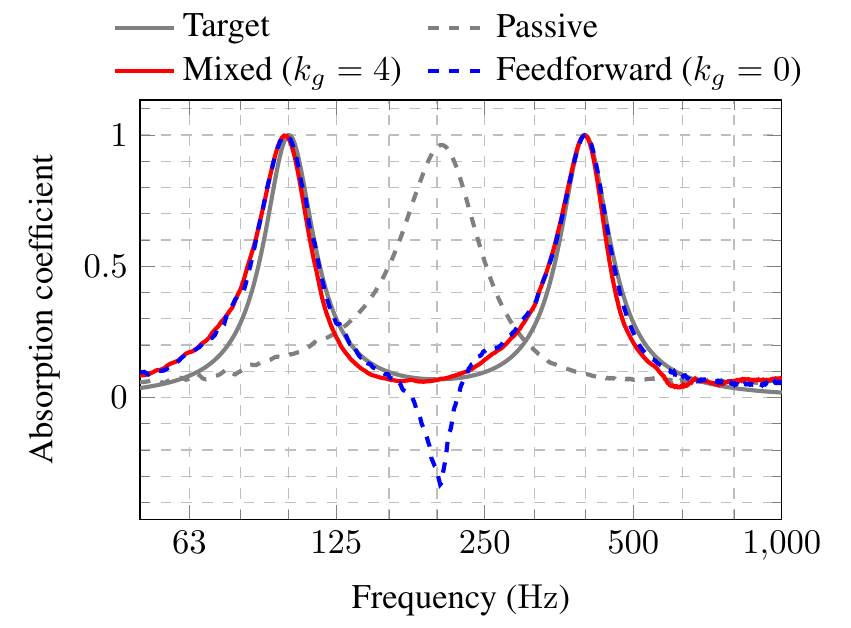}
    \caption{Experimentally obtained absorption coefficients for the two-degree-of-freedom absorber, with $\hat{F} = 0.95F$}
    \label{fig:experimental_2DOF_alpha}
\end{figure}

The three considered target impedances are described by the parameters from Table \ref{tbl:ctrl_parameters}.
To highlight the advantage of the mixed feedforward-feedback controller, a 5\% error was purposely included in the model of the loudspeaker, needed to build the controller transfer functions, such that $\hat{F} = 0.95F$.
In Figure \ref{fig:experimental_1DOF_alpha}, Figure \ref{fig:experimental_broadband_alpha} and Figure \ref{fig:experimental_2DOF_alpha}, the passive, the target and the achieved absorption coefficients with and without the feedback contribution are drawn.

Like for the numerical study, it is observed that the passive resonant behaviour is still present in the achieved impedances without feedback, reaching in some cases a negative value of absorption and adding a degree of freedom to the achieved impedance.
The mixed feedforward-feedback controller is capable to overcome this issue, does truly behave as the target and is more accurate, especially around the passive resonance of the loudspeaker.
Note that the lack of precision at lower frequencies (i.e., lower than \SI{100}{\hertz}) for both controllers is inherent to the Kundt's tube measurement.
Indeed, the termination reflection coefficient $\Gamma(s)$ is given in \cite{ISO_10534_2} as
\begin{equation}
    \Gamma(s) = \frac{H_{12}(s) - e^{-jk\Delta x}}{e^{jk\Delta x} - H_{12}}e^{-2jkx1} ,
    \label{eq:kundt_gamma}
\end{equation}
where $H_{12}(s) = p_2(s)/p_1(s)$ is the transfer function between the two measurement microphones, $k$ is the wave number and $\Delta x$ and $x_1$ are dimensions visible in Figure \ref{fig:experimental_setup}.
When the frequency tends to zero, equation \eqref{eq:kundt_gamma} becomes ill conditioned because both $H_{12}(s)$ and $e^{\pm jk\Delta x}$ tend to one.
Equation \eqref{eq:kundt_gamma} is therefore very sensitive to the measurement errors in $H_{12}$ for low frequencies.

\section{Conclusions}
\label{sec:05_conclusion}
This article presented a new direct impedance control architecture providing a more accurate and robust control on the actual impedance than previously reported in the literature.
The concept of mixed feedforward-feedback control is based on an already existing feedforward implementation, but to achieve a better accuracy, it is combined with a feedback loop that relies on the sensing of the displacement of the actuator to adjust the driving current.
Displacement sensing is done through a microphone placed in the enclosure of the loudspeaker, effective at low frequencies.
Even if it is not a noticeable improvement for broadband absorption, as targeted by the feedforward architecture \cite{Rivet17}, it does significantly improve the passivity, and thus the stability, of a multi-degree-of-freedom absorber, as formerly used in aircraft engine noise reduction applications.
Additionally, in such an environment, the estimated parameters of the absorber might change significantly with the static pressure, surrounding temperature or humidity.
With the feedback contribution, the sensitivity to errors is lowered, and is therefore more adapted to drifting parameters.

This design could be further improved, typically by investigating different relations between $H_1$ and $H_2$ in equation \eqref{eq:H1_H2_link}.
Also, a more sophisticated model of the relationship between the membrane velocity and the pressure in the cavity could be considered to extend the feedback contribution to higher frequencies or larger loudspeaker enclosures.
For this, a more elaborated fitting should be used in equation \eqref{eq:cb_estimate} rather than a constant real value.
Furthermore, the mixed feedforward-feedback control could also be used to linearize actuators at high sound pressure levels, at which their stiffness is no longer linear and typically depends on the membrane position.

\appendix
\section{Current source}
\label{sec:06_current_source}
\begin{figure}
    \centering
    \includegraphics{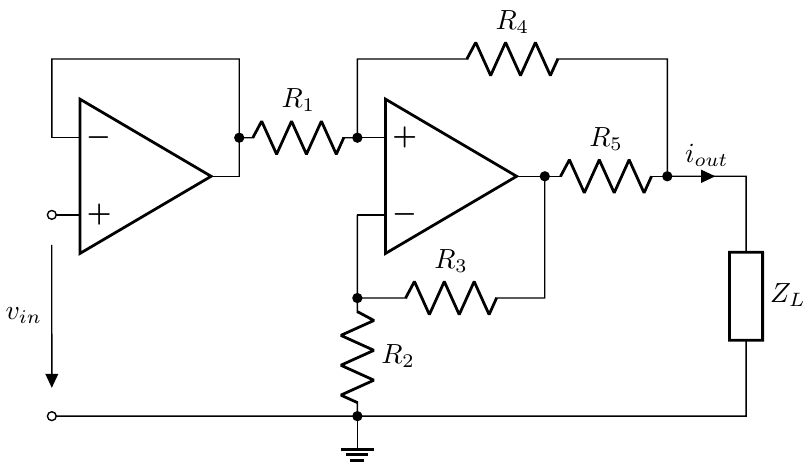}
    \caption{Voltage controlled current source schematic. $R_1=R_2=\SI{92}{\kilo\ohm}$, $R_3=R_4=\SI{1.1}{\kilo\ohm}$ and $R_5=\SI{1.2}{\ohm}$.}
    \label{fig:current_source}
\end{figure}

The voltage controlled current source used to drive the loudspeaker for the experimental measurements is depicted in Figure \ref{fig:current_source} and is inspired from the application report \cite{AN1515}.
The chosen operational amplifier is a TL288CP from Texas Instruments.
The output current can be shown to be
\begin{equation}
    \begin{split}
        i_{out} = &v_{in}\frac{R_3R_4 + R_2(R_4 + R_5)}{(R_1+R_4)R_2R_5} \\
        & + v_{out}\frac{R_1R_3 - R_2(R_4 + R_5)}{(R_1+R_4)R_2R_5} .
    \end{split}
\end{equation}
When $R_1=R_2$ and $R_3=R_4 + R_5$, it simplifies to a proportional relation between input voltage and output current, regardless of the load impedance $Z_L$:
\begin{equation}
    i_{out} = v_{in}\frac{R_3}{R_1R_5} .
\end{equation}
With the values from Figure \ref{fig:current_source}, a suitable voltage controlled current source for driving a loudspeaker is obtained:
\begin{equation}
    i_{out} = v_{in}\cdot\SI{9.97}{\milli\ampere\per\volt} - v_{out}\cdot\SI{10.7}{\micro\ampere\per\volt} .
\end{equation}

The current delivered by the operational amplifier is
\begin{equation}
    i_{oa} = i_{out} \left(\frac{R_3-R_5}{R_3}\,\frac{R_1+R_3+R_5}{R_1+R_3-R_5} + \frac{2Z_L}{R_1+R_3-R_5} \right) ,
\end{equation}
which is approximately the output current of the current pump since $R_5$ and the $Z_L$ are both much smaller than $R_3$.
For the 2 DOF case from Table \ref{tbl:ctrl_parameters}, the highest current is required when all the incident pressure is concentrated at \SI{100.8}{\hertz}.
The maximal output current for the TL288CP is of \SI{80}{\milli\ampere}, which is reached when the incident pressure is \SI{117}{\decibel} SPL at \SI{100.8}{\hertz}.

\section{Microphone position in the cavity}
\label{sec:06_microphone_placement}
For wavelengths much smaller than the dimension of the enclosure of the loudspeaker, the pressure in the cavity is proportional to the displacement of the membrane.
However, as the frequency increases, the model of the box is becoming worse, and cavity modes appear.
The position of the microphone in the cavity can help mitigate this effect.

\begin{figure}
    \centering
    \includegraphics{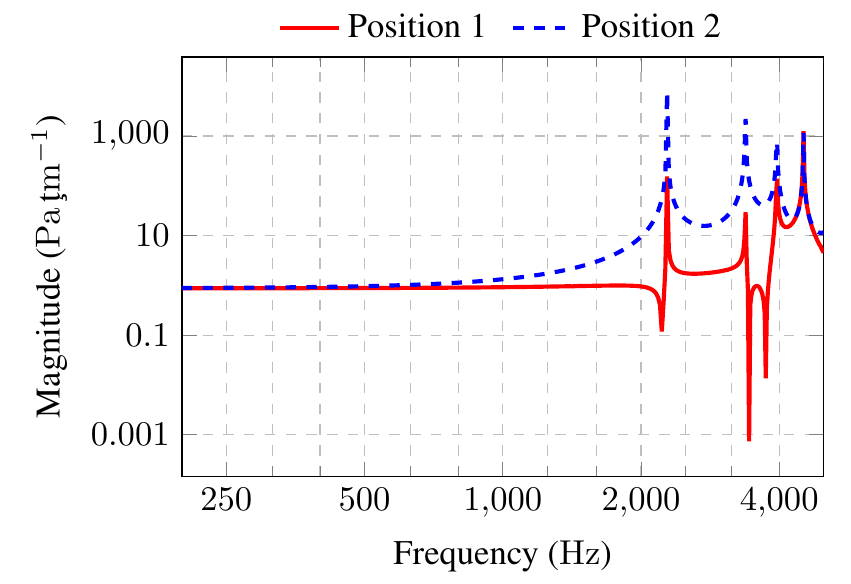}
    \caption{Simulated transfer function between rear microphone pressure and membrane displacement $C_{sb}(s)=\xi(s)/p_b(s)$}
    \label{fig:comsol_Csb}
\end{figure}

\begin{figure}
    \centering
    \includegraphics{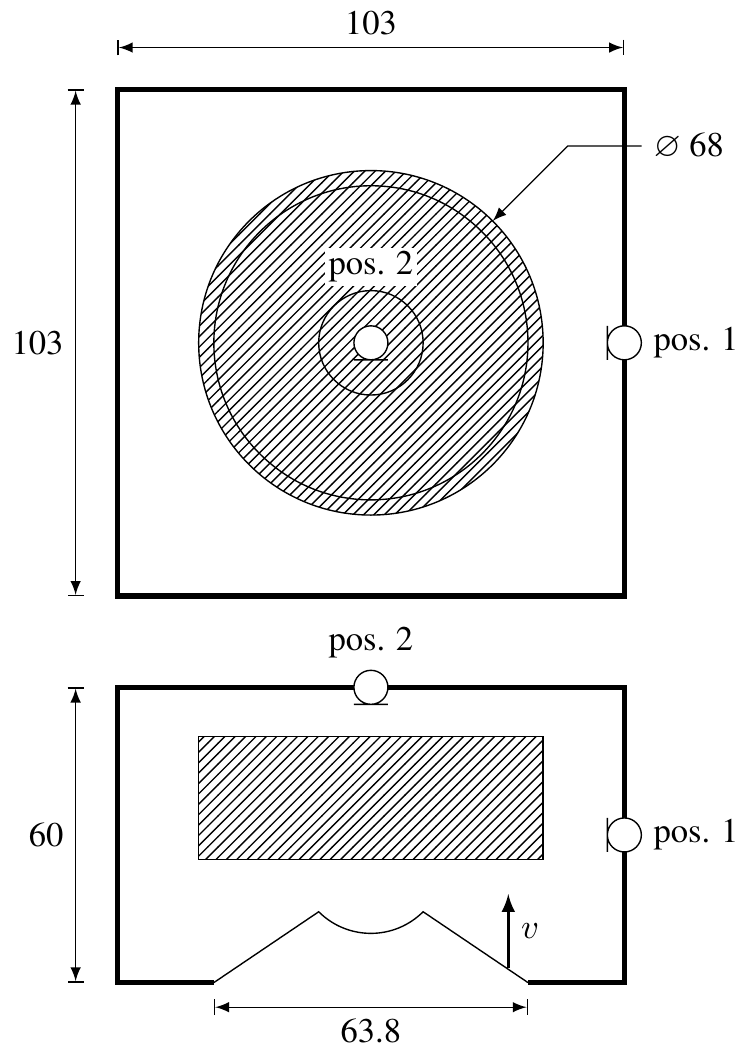}
    \caption{Simulated geometry, with the two microphones positions. Membrane is drawn in a thin line and the magnet is hatched. Units in \si{\milli\meter}}
    \label{fig:comsol_geom}
\end{figure}

Frequency-domain simulations have been conducted using the finite element simulation software COMSOL Multiphysics to find an optimal microphone position.
The obtained relationships from the membrane displacement to the pressure at the position of the microphone $p_b/\xi$ are reported in Figure \ref{fig:comsol_Csb} for the two geometries shown in Figure \ref{fig:comsol_geom}.
In this graph, it is visible that the first cavity mode happens at \SI{2.2}{\kilo\hertz}.
The response of the microphone at position 1 has the flattest response up to this frequency and is therefore chosen in the experimental absorber prototype.
However, to avoid instabilities at high frequencies, some melamine foam was added in the enclosure, which will damp higher frequencies and remove the undesired spikes.

\section*{Acknowledgment}
This project has received funding from the Clean Sky 2 Joint Undertaking under the European Union's Horizon 2020 research and innovation program under grant agreement No 821093.

This publication reflects only the authors' view, and the JU is not responsible for any use that may be made of the information it contains.

\bibliographystyle{ieeetr}
\bibliography{references}

\end{document}